\documentclass[aip, amsmath,amssymb,reprint]{revtex4-2}

\usepackage{graphicx}
\usepackage{dcolumn}
\usepackage{bm}
\usepackage{siunitx}
\usepackage{hyperref}
\hypersetup{colorlinks = true,linkcolor=blue,
     filecolor=blue,
     urlcolor=blue,
     citecolor = blue, pdfauthor=author}

\begin{document}

\preprint{AIP/123-QED}

\title{The emergence of low-frequency dual Fano resonances in chiral twisting metamaterials}

\author{Brahim Lemkalli}
\email{\textcolor{black}{brahim.lemkalli@edu.umi.ac.ma}}
\affiliation{Laboratory for the Study of Advanced Materials and Applications, Department of Physics, Moulay Ismail University, B.P. 11201, Zitoune, Meknes, Morocco}

\author{Muamer Kadic}
\affiliation{Institut FEMTO-ST, UMR 6174, CNRS, Universit\'{e} de Bourgogne Franche-Comt\'{e}, 25000 Besan\c{c}on, France}

\author{Youssef El Badri}
\affiliation{Laboratory of optics, information processing, Mechanics, Energetics and Electronics, Department of Physics, Moulay Ismail University, B.P. 11201, Zitoune, Meknes, Morocco}

\author{S\'{e}bastien Guenneau}
\affiliation{UMI 2004 Abraham de Moivre-CNRS, Imperial College London, SW7 2AZ, UK}

\author{Abdellah Mir}
\affiliation{Laboratory for the Study of Advanced Materials and Applications, Department of Physics, Moulay Ismail University, B.P. 11201, Zitoune, Meknes, Morocco}

\author{Younes Achaoui}
\affiliation{Laboratory for the Study of Advanced Materials and Applications, Department of Physics, Moulay Ismail University, B.P. 11201, Zitoune, Meknes, Morocco}
\date{\today}

\begin{abstract}
In the current work, through a finite element analysis, we demonstrate that a configuration of chiral cells having syndiotactic symmetry provides dual Fano resonances at low frequency. From the phononic dispersion and transmission response, we compare the signature provided by a composite made of chiral cells to the ones of homogeneous medium, isotactic nonchiral, and isotactic chiral beams. The study results in an innovative design of a mechanical metamaterial that induces the Fano resonance at low frequency with a relatively high quality factor. This might be a significant step forward for mechanical wave filtering and detection. Performances have been evaluated using a sensor that will be implemented as a thermometer.
\end{abstract}


\maketitle
\section{Introduction} \label{sec1}
In recent years, the emergence of composite-structured materials has heralded significant advancements in mechanical engineering \cite{dalela2022review}. As a result, new generations of man-made materials, known as "metamaterials," are created, allowing mechanical behaviors to be adapted with new characteristics beyond the intrinsically well known \cite{xiao2020active}. From an elastodynamic viewpoint, these allow the manipulation and control of acoustic wave propagation by both mechanisms, namely local resonance and Bragg scattering \cite{achaoui2011experimental, kadic2013metamaterials}. Besides, mechanical metamaterials are well known by their various exotic parameters in the static regime, including negative Poisson's ratio \cite{lakes2017negative}, flexibility \cite{bertoldi2017flexible}, and twist conversion \cite{frenzel2017three, zhong2019novel}, which leads to a "dynamic paradigm," used today in a wide range of applications. For instance, auxetic metamaterials were proposed in order to enhance seismic shielding against surface waves \cite{ungureanu2015auxetic}. Besides, metamaterials with a twist can exhibit a distinct feature called acoustic activity. This converts the linear polarization of a transverse wave to circular polarization \cite{frenzel2019ultrasound}. Recently, twisting metamaterials demonstrated the conversion of longitudinal waves into twist waves \cite{lemkalli2022longitudinal, xu2022origami}.

In general, the local resonance is caused by the coupling between a discrete resonance and a continuous state, which causes the appearance of a peak at the resonance frequency followed or preceded by the dip of the anti-resonance. This mechanism is a consequence of constructive and destructive interferences, respectively, previously reported in the field of optics \cite{fano1961effects}. Since its discovery more than 60 years ago \cite{fano1961effects}, the prominent Fano resonance has piqued the interest of scientists due to its asymmetric nature, which is used in some relevant applications \cite{zhou2014progress} such as filtering \cite{shuai2013double} and detection \cite{luk2010fano}.

As a mechanical counterpart, this sort of resonance has gained prominence \cite{wang2020robust}. Several devices based on mechanical Fano resonance have been developed in recent years \cite{el2008transmission}, including concentrated pipes \cite{amin2015acoustically}, Helmholtz resonators \cite{qi2014interference}, and phononic crystals \cite{zaki2020fano, goffaux2002evidence,oudich2018rayleigh}. However, the dimensions of these structures, notably phononic crystals, are equivalent to or even larger than the wavelengths; also, the Fano resonance effect occurs in just one operational frequency range. Multi-band systems with sub-wavelength dimensions and a high quality factor at low frequencies remain a major challenge for the development of multi-band and multi-functional devices \cite{zaki2020fano}. Dual Fano resonators for low frequencies have recently been developed, employing an array of units made up of two types of cell units containing multiple cavities, each with its own specific set of characteristics \cite{sun2019dual}. These are based on the emergence of acoustic metamaterials\cite{cummer2016controlling} with dimensions smaller than the wavelength, leading to exceptional elastic wave manipulation abilities.

\begin{figure*}
    \centering
    \includegraphics[width=13cm,angle=0]{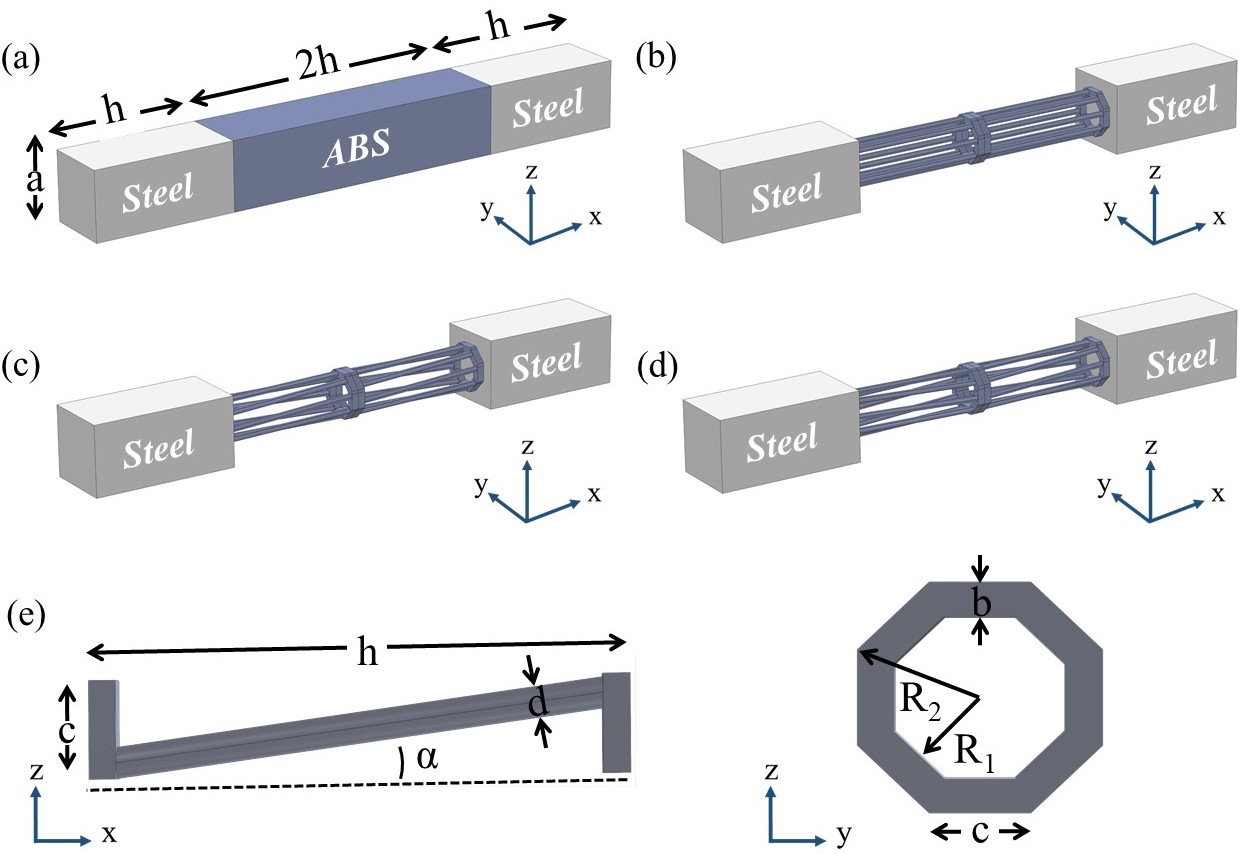}
    \caption{Schematics of the beams. (a) The homogenous medium cell. (b) The nonchiral isotactic cell ($\alpha=0$). (c) The chiral isotactic cell $\alpha=arctn(\frac{h}{c})$. (d) The chiral syndiotactic cell $\alpha=arctn(\frac{h}{c})$. (e) The geometrical parameters of the cells, the two octagonal plates are separated by a distance of $h=30$\si{mm} by rods with diameter of $d=1.2$\si{mm} inclined by an angle $\alpha$ equal to $arctn(\frac{h}{c})$ with the side  of the octagon equal to $c=3.9$\si{mm} and the radii $R_1=5.08$\si{mm} and $R_2=3.3$\si{mm} and $b=1.4$\si{mm}. The beams have a width of $a=14.4$\si{mm} and length of $4h$.}
    \label{Figure 1}
\end{figure*}

In this study, we leverage the design of a metamaterial with a twist to generate double Fano resonance at low frequency, inspired by the chiral tacticity in metamaterials \cite{bergamini2019tacticity}. In Section \ref{sec2}, we demonstrate numerically that a chiral syndiotactic cell generates local resonance. By connecting two cells in such a way that the contact plane between the two cells forms a mirror, the Fano resonance fingerprint is the direct consequence of the coupling between a longitudinal continuum and the discrete state of the chiral unit-cell. In Section \ref{sec3}, we propose an application of the dual Fano resonances to detect temperature changes in water.

\section{Elastodynamic Characteristics}\label{sec2}

In this section, we analyzed the elastodynamic behavior of four structures in order to demonstrate the presence of the Fano resonance. These structures have rectangular beams with a length of $4h$ and a width of $a$ made of two media; one homogeneous in steel at the beam borders with a length of $h$ and the other inside, which is a cell in Acrylonitrile Butadiene Styrene (ABS) with a length of $2h$ alternating the four unit cells. The first cell is purely a homogeneous medium (\autoref{Figure 1}(a)). The second is a nonchiral isotactic unit cell throughout two cells composed of non-inclined rods ($\alpha=0$) connected to octagonal plates (\autoref{Figure 1}(b)). The third is a chiral isotactic unit cell in which two cells composed of rods inclined by ($\alpha=atan(h/b)$) are connected to octagonal plates (\autoref{Figure 1}(c)). The fourth cell is a chiral syndiotactic cell composed of two chiral cells with inclined rods ($\alpha=atan(h/b)$) attached to octagonal plates (\autoref{Figure 1}(d)). These two cells are connected by a plane of symmetry mirror. 

To determine the elastodynamic behaviors of the four beams, we used the commercial software COMSOL Multiphysics to solve the Navier equations in the weak form. We considered all the materials used in the simulations as isotropic linear elastic materials. These are depicted in \autoref{Table1}.
\begin{table}[h]
    \centering
    \caption{ \label{Table1}The materials parameters}
     \begin{ruledtabular}
    \begin{tabular}{c c c c}
     Materials & Young's modulus & Poisson's ratio & Density   \\
      & (\si{GPa}) & & (\si{kg/m^3})\\\hline\\
     Steel & 201 & 0.33 & 7843\\ 
     ABS & 2.6 & 0.4 & 1020\\
    \end{tabular}
     \end{ruledtabular}
\end{table} 

\begin{figure*}
    \centering
    \includegraphics[width=14cm,angle=0]{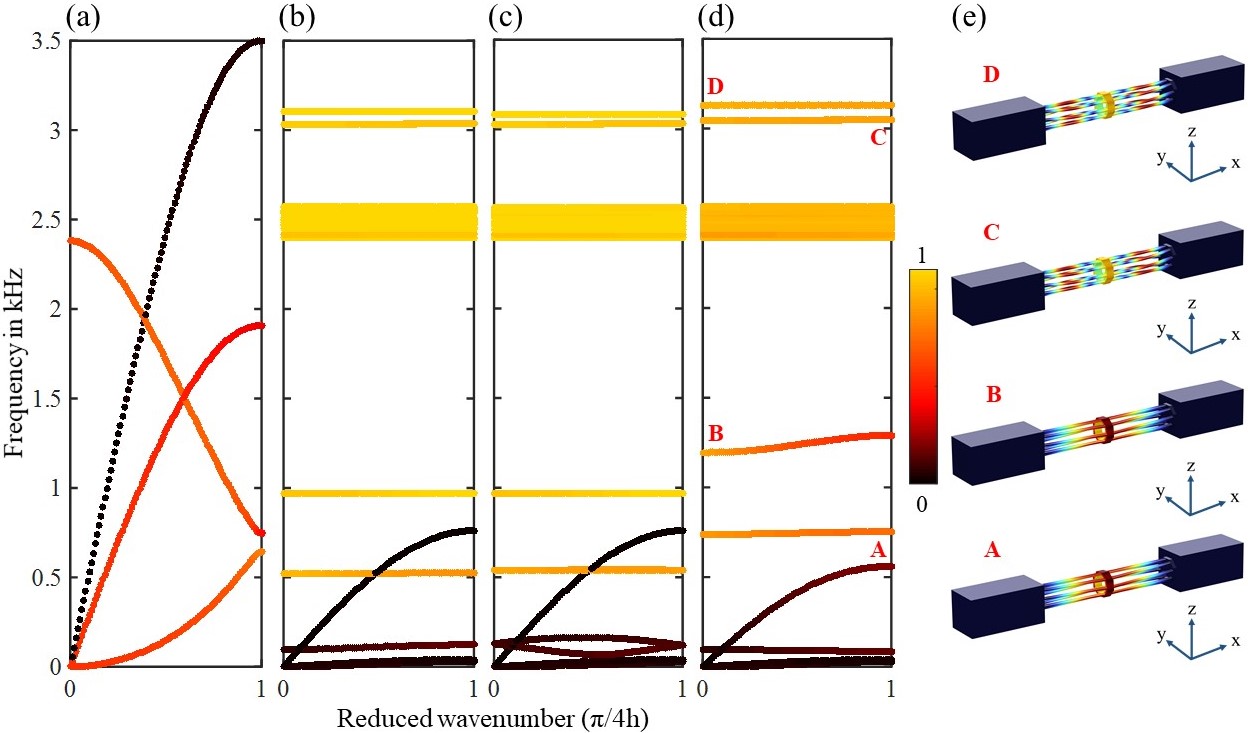}
    \caption{The phononic dispersion curves along the $x$-direction in the first Brillouin zone ($\Gamma X$) for the four beams. (a) The homogeneous medium cell. (b) The nonchiral isotactic cell. (c) The chiral isotactic cell. (d) The chiral syndiotactic cell. (e) Screenshots of the syndiotactic chiral cell beam's eigenmodes at points $A$, $B$, $C$, and $D$.}
    \label{Figure 2}
\end{figure*}
The first step was to calculate the phononic dispersion curves towards the $x$-direction and analyze the eigenmodes of the four beams. To elucidate the mechanisms that govern the interaction of localised modes (flat modes) with longitudinal modes, which gives rise to the local resonance. We calculated mode polarization using equation \ref{Eq01}, which is represented by the color bar in the dispersion curves, as depicted in \autoref{Figure 2}.
\begin{equation}\label{Eq01}
    p_{yz}=\frac{\iiint \sqrt{|u_y|^2+|u_z |^2}  dV_1}{\iiint \sqrt{|u_x |^2+|u_y |^2+|u_z |^2 )}  dV_{tot} },
\end{equation}
where $V_1$ is the volume of the inner cell and $V_{tot}$ is the total volume of the beam. 

The beam with the cell of homogeneous medium (ABS) (\autoref{Figure 2}(a)) exhibits four fundamental modes: bending and transverse, which are degenerated in the present case because of the symmetry in the $yz$-plane, plus the other two modes: twisting and longitudinal. As illustrated, the first three modes have their polarization active in the $yz$-plane, with the exception of the longitudinal mode, which remains fully polarized all along the $x$-direction.
However, when we substitute the homogeneous medium (ABS) with the nonchiral isotactic cell (\autoref{Figure 2}(b)) in the beam with a length of $2h$, the first two modes remain degenerate but have shifted down towards low frequencies. The polarization indicates that the flat modes do not interfere with the longitudinal mode. 

Analogously, the longitudinal mode remains polarized along the $x$-direction in the isotactic chiral cell (\autoref{Figure 2}(c)), regardless of the fact that the first two modes have undergone degeneracy lifting (the transverse modes travel with different velocities in the first Brillouin zone ($\Gamma X$)) as a consequence of the symmetry in the $yz$-plane. In other words, the effect of the isotactic chiral cell has no influence on the polarization of the longitudinal mode (the coupling between the flat modes does not take place with the longitudinal mode). 

However, due to the presence of a distinct symmetry plane inside the syndiotactic cell, the first two modes do not undergo degeneracy lifting (\autoref{Figure 2}(d)). On the other hand, there is interference between a localized mode, polarized in the $yz$-plane, and the longitudinal mode, which causes the local resonance composed of two resonances, symmetric and anti-symmetric, which is produced as a result of the coupling of the flat mode with the longitudinal mode, resulting in the presence of the Fano resonance near $1$ \si{kHz}, as indicated by the red color of the mode polarization around this frequency. 

To illustrate that the interference between the localized twist mode and the longitudinal is entirely responsible for the appearance of the local resonances, screenshots of the modes at the local resonance are displayed in \autoref{Figure 2}(e). $A$ has the coordinates of ($k$, $\omega$)=($1$, $862.8)$ and $B$($0$, $1138$). According to the dispersion curve, these two points represent the initial local resonance. Both images show a localized displacement in the center of the syndiotactic chiral cell. These images suggest that at these frequencies, the effect of longitudinal-twist conversion between the two cells is active, which results in local resonances. Around $3$ \si{kHz} of \autoref{Figure 2}(d), we can discern two polarized modes in the $yz$-plane, indicating the resonance pattern that forms the Fano resonance. In those, we consider two points $C$ and $D$, as seen in the screenshots in \autoref{Figure 2}(e). This last Fano resonance is produced by a high-order twist; the total displacement is localized in three regions of the syndiotactic chiral cell: in the center and in the middle of the rods, as seen in the \autoref{Figure 2}(e).


After demonstrating the existence of the Fano resonance in the syndiotactic chiral cell using eigenvalue analysis, we have investigated the transmission analysis of a longitudinal wave through these four beams. We added free media in steel and Perfectly Matched Layers at each extremity. We used equation $2$ to calculate the longitudinal wave transmission along the $x$-direction for the four beams. 
\begin{equation}\label{Eq02}
    T=20\times log_{10}\frac{\iiint |u_x|^2  dV_{output}}{\iiint |u_x|^2  dV_{input}},
\end{equation}
\begin{figure}[h]
    \centering
    \includegraphics[width=8cm,angle=0]{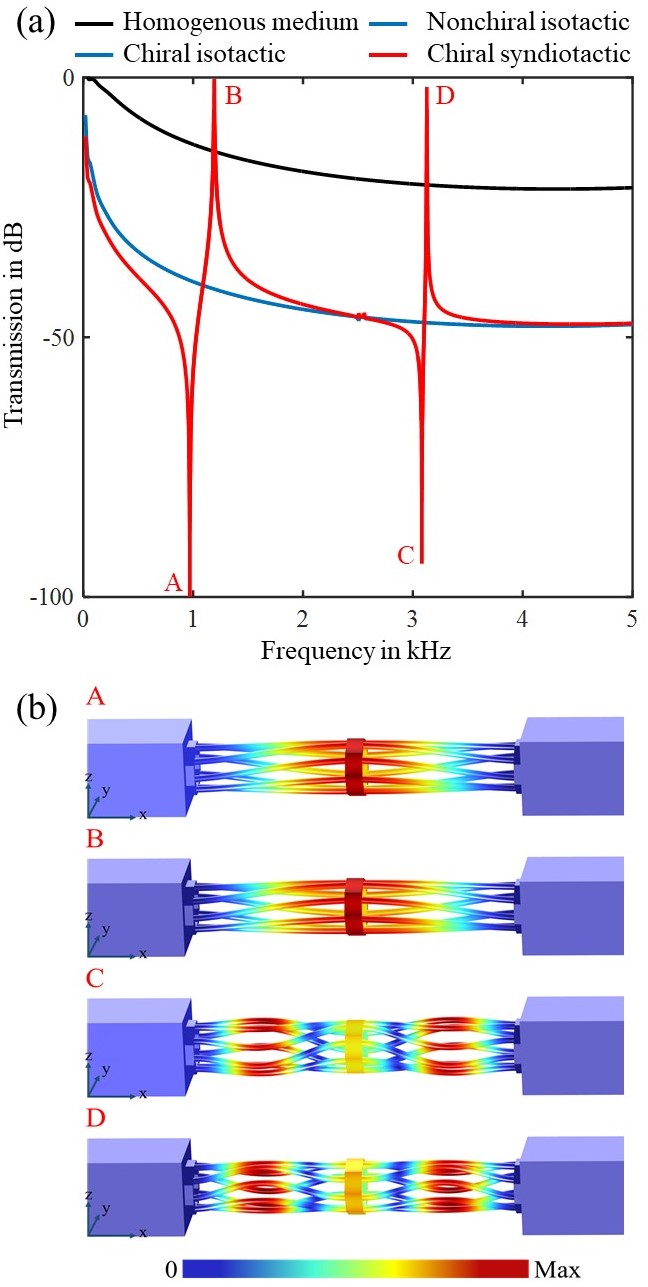}
    \caption{(a) Transmission spectrum of the homogeneous medium cell in black color. The nonchiral isotactic cell in blue color. The chiral isotactic cell in blue color. The chiral syndiotactic cell in red color. (b) Screenshots of the syndiotactic chiral beam in the anti-resonance and resonance peaks, at point A of frequency of $970$ \si{Hz}, at point B of frequency of $1192$ \si{Hz}, at point C of frequency of $3082$ \si{Hz}, and at point D of frequency of $3127$ \si{Hz}.}
    \label{Figure 3}
\end{figure}

\autoref{Figure 3}(a) depicts the transmission curves of the beams with a homogeneous medium cell, both isotactic nonchiral and chiral cells, and a syndiotactic chiral cell, as shown by the black, blue, and red curves, respectively. The first three beams have no Fano resonances, whereas the fourth one contains two Fano resonances. This indicates that the syndiotactic chiral structure exhibits a dual Fano resonances at low frequencies. We evaluated the quality factor, which corresponds to the ratio of the resonance frequency to the full width frequency at half-maximum of each peak. The first peak at the resonance frequency of $1.191$ \si{kHz} has a Q-factor of $350$, while the second occurs at the resonance frequency of $3.127$ \si{kHz} has a Q-factor of $11,010$. \autoref{Figure 3}(b) shows screenshots at resonance and anti-resonance frequencies, which are the peak and the dip for the two Fano resonances that exist exclusively in the syndiotactic chiral cell. The first resonance represents the first-order twist, as indicated by the dispersion curves, while the second resonance is the high-order twist.

\section{Liquid sensing application}\label{sec3}

\begin{figure}[h] 
    \centering
    \includegraphics[width=8cm,angle=0]{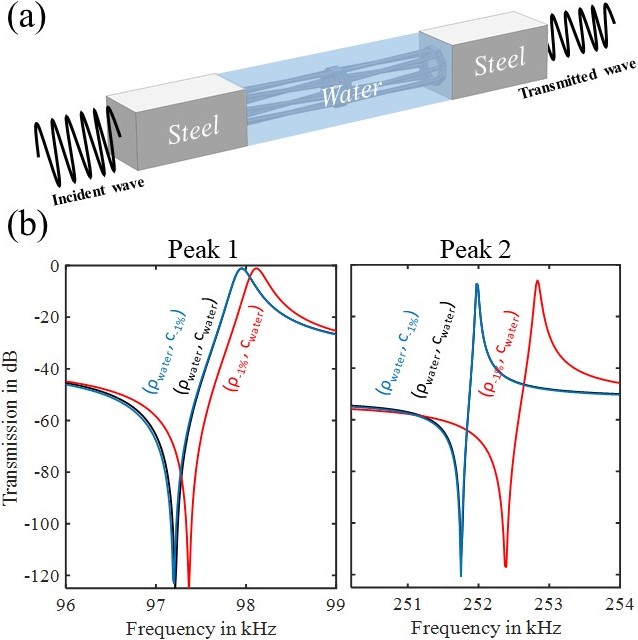}
    \caption{Chiral syndiotactic beam as liquid sensor.  (a) The sensor's design. (b) Longitudinal transmission for both peaks as a function of $-1\%$ density and speed of sound variation.}
    \label{Figure 4}
\end{figure}

Phononic crystals and metamaterials have gained significant attention in the realm of sensing, particularly as an innovative resonant platform for analyzing liquid properties. The general idea is based on the incorporation of liquid as a constituent in the phononic crystal or within a cavity localized in the perfect structure \cite{lucklum2012two, ke2011sub, oseev2018study}. Their sensing functionality, in particular, is realized through the solid-liquid interaction \cite{wang2017tunable, wang2022reconfigurable}. Based on this approach, we propose a sensor based on the chiral syndiotactic cell, as shown in \autoref{Figure 4}(a), which is a beam composed of two homogeneous media in steel and a syndiotactic chiral cell in ABS submerged in water with a dimension two order of magnitude smaller than the geometrical parameters outlined in Section \ref{sec2}. This makes the sensor operates at frequencies around $100$ \si{kHz}, which is low frequency in comparison to the sensors described in the literature.

In order to evaluate the potentiel of the presented sensor to detect changes in liquid properties, we defined the sensitivity using equation \ref{03}, which quantifies the frequency shift of each resonance peak in response to a slight change in the liquid properties. Additionally, we use equation \ref{04} to define the figure of merit (FoM), which evaluates if two nearly identical media can be distinguished.

\begin{equation}\label{03}
    S_i=\frac{\Delta f_i}{\Delta T},
\end{equation}
\begin{equation}\label{04}
    FoM_i=\frac{S_i \times Q_i}{f_i},
\end{equation}
where $i$ symbolizes peak $1$ or peak $2$, $T$ is the temperature, $Q_i$ is the quality factor of each resonance peak, and $f_i$ represents the resonance frequency for each peak.

In the first step, we assessed the syndiotactic beam sensor's capabilities employing investigations given in recent works using phononic crystals, in which they varied the water density and sound velocity by $1\%$ and determined the characteristic parameters of each variation \cite{gueddida2021tubular, gueddida2022acoustic}. The longitudinal transmission responses are depicted in \autoref{Figure 4}(b). As shown in \autoref{Figure 4}(b), the syndiotactic sensor is not highly sensitive to the variations of the sound speed, but is sensitive to variations in density. We summarize the calculated parameters of density variation for the two peaks in \autoref{Table 2}. 

\begin{table}[h!]
\caption{\label{Table 2} Frequency, $Q$-factor, sensitivity, and figure of merit of the two peaks for $-1\%$ of density water variation.}
\begin{ruledtabular}
\begin{tabular}{c c c}
                 & Peak 1 & Peak 2\\\hline
         Frequency (\si{kHz})& 97.9& 251.9\\
         Q & 840 &$ 17 \times 10^3$\\
         $S_{-1\%\rho}$ (\si{Hz/{kg m^{-3}}}) & 16 & 100\\
         $FoM_{-1\%\rho}$ (\si{1/{kg m^{-3}}})& 0.14& 6.74\\ 
    \end{tabular}
\end{ruledtabular}
\end{table}

The first peak at $97.9$ \si{kHz} has a $Q$-factor of $840$ and a sensitivity to density variation of $1\%$ equal to $16$ \si{Hz/{kgm^{-3}}} and a figure of merit of $0.14$ \si{1/{kg m^{-3}}}. Regarding the second peak around $251$ \si{kHz}, a $Q$-factor of $17,000$, a sensitivity of $100$ \si{Hz/{kg m^{-3}}}, and a figure of merit of $6.74$ \si{1/{kg m^{-3}}}. The parameters obtained for the $1\%$ density variation are comparable to those published in the literature \cite{gueddida2021tubular, gueddida2022acoustic}. These parameters show that at low frequencies, the chiral syndiotactic cell has an interesting feature in the detection of density variation unlike the sound speed.

\begin{figure}
    \centering
    \includegraphics[width=8.5cm,angle=0]{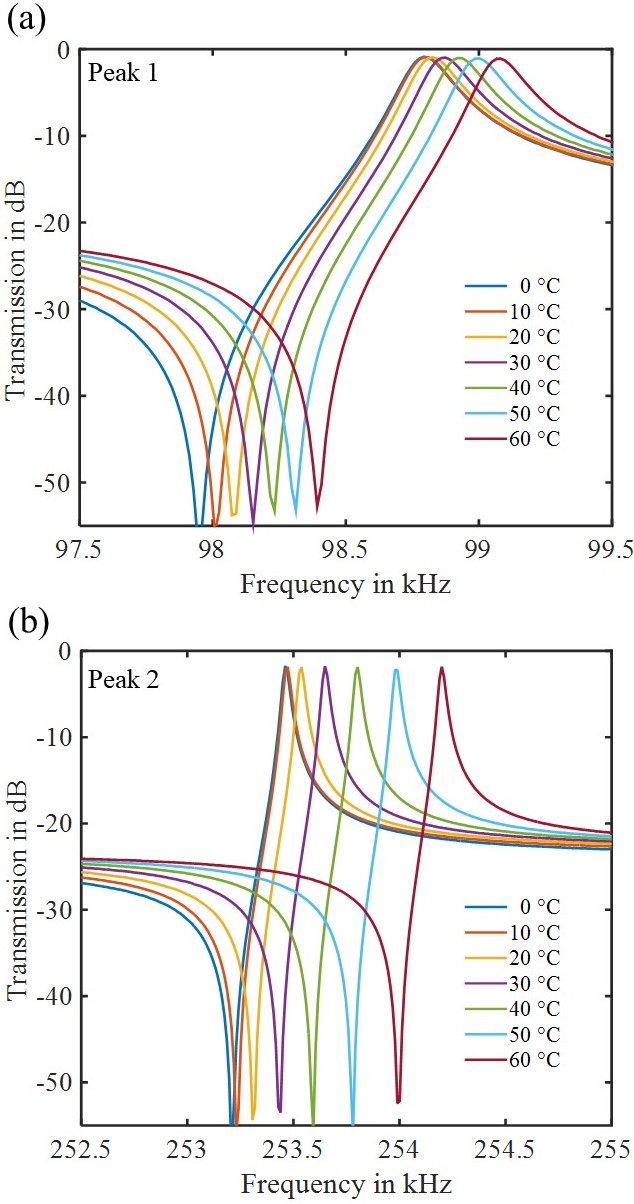}
    \caption{The evolution of longitudinal transmission as a function of water temperature variation. (a) The first peak. (b) The second Peak.}
    \label{Figure 5}
\end{figure}

In the second step, we used the syndiotactic chiral beam to detect changes in water temperature, in which the density and the speed of sound depend on the temperature of water, as depicted in \autoref{Table 3}. We computed the longitudinal transmission in function of water temperature variation, as illustrated in \autoref{Figure 5}. The frequencies, quality factors, sensitivities, and FoMs of the two peaks are all described in \autoref{Table 4}. The sensitivity of each peak was determined by estimating two successive values of the frequency shift produced by the temperature change.

\begin{table} 
    \caption{ \label{Table 3} The density and speed of sound as a function of water temperature variation.}
    \begin{center}
     \begin{ruledtabular}
    \begin{tabular}{c c c}
    Temperature  &  Density  & Speed of sound\\
    (\si{\degree C})& (\si{kg/m^3})& (\si{m/s})\\ \hline
    0&999&1403\\ 10&999&1447\\ 20&998&1481\\30&995&1507\\
    40&992&1526\\ 50&988&1541\\ 60&983&1541\\
    \end{tabular}
    \end{ruledtabular}
    \end{center}
\end{table}

\begin{table}[h]
\caption{\label{Table 4} Frequency, $Q$-factor, sensitivity, and figure of merit of the two peaks for temperature variation in water.}
\begin{ruledtabular}
\begin{tabular}{c c c}
          &Peak 1 & Peak 2  \\\hline\\
         Frequency (\si{kHz}) &  98.8 & 253.5  \\
          $Q$-factor  & 840 & $ 17 \times 10^3$\\
         Sensitivity (\si{Hz/{\degree C}}) & 3 & 20\\
         FoM (\si{1/{\degree C}}) & 0.02 & 1.34\\
         \\
    \end{tabular}
\end{ruledtabular}
\end{table}

As the water temperature increases, the frequency of the two Fano resonance peak shifts. Thus, the characteristic frequencies of the two Fano resonance peaks are sensitive to both the speed of sound and density of water as a function of temperature. In other words, the chiral syndiotactic beam can exhibit two peaks that can change the resonance frequency based on the liquid's material properties, as illustrated in \autoref{Figure 5}. The first peak has lower sensitivity than the second peak; however, both peaks have acceptable characteristics in terms of increasing dimension and using a low frequency when compared to the phononic sensor, which uses a frequency of hundreds of \si{GHz}. 

The findings suggest that the syndiotactic chiral beam could be used as a temperature sensor. Due to the geometric size, two peaks occur at low frequencies, which indicates that their characteristics are indeed very small. It should be noted that in order to achieve the highest sensitivity value, it must use a structure with geometrical parameters scaled by 0.1 compared to the presented configuration. This means that both frequency response and sensitivity will be multiplied by a factor of $10$.

\section{Conclusion}

To conclude, we used the finite element method to calculate phononic dispersion and transmission. This was done in order to demonstrate the presence of dual Fano resonances in chiral metamaterials with twist. Furthermore, we proved that the beam with the chiral syndiotactic cell based on the two octagonal plates exhibited dual Fano resonances at low frequencies, one at $1$ \si{kHz} and the other at $3$ \si{kHz}. This characteristic has been compared to other beams with a homogeneous medium, an isotactic nonchiral cell, and an isotactic chiral cell. The interference of localized twisting and longitudinal modes, in particular, causes low-frequency local resonances, also known as Fano resonances. Following that, the presence of dual Fano resonances in syndiotactic beam metamaterials is used to detect liquid properties such as water density and sound speed. Finally, this study demonstrated that syndiotactic beam metamaterials with twist can be used as temperature sensors, exhibiting considerable sensitivity and quality factors for the proposed size and for low frequencies.\\


\bibliography{mybibfile}

\providecommand{\noopsort}[1]{}\providecommand{\singleletter}[1]{#1}%
\begin{thebibliography}{33}%
\makeatletter
\providecommand \@ifxundefined [1]{%
 \@ifx{#1\undefined}
}%
\providecommand \@ifnum [1]{%
 \ifnum #1\expandafter \@firstoftwo
 \else \expandafter \@secondoftwo
 \fi
}%
\providecommand \@ifx [1]{%
 \ifx #1\expandafter \@firstoftwo
 \else \expandafter \@secondoftwo
 \fi
}%
\providecommand \natexlab [1]{#1}%
\providecommand \enquote  [1]{``#1''}%
\providecommand \bibnamefont  [1]{#1}%
\providecommand \bibfnamefont [1]{#1}%
\providecommand \citenamefont [1]{#1}%
\providecommand \href@noop [0]{\@secondoftwo}%
\providecommand \href [0]{\begingroup \@sanitize@url \@href}%
\providecommand \@href[1]{\@@startlink{#1}\@@href}%
\providecommand \@@href[1]{\endgroup#1\@@endlink}%
\providecommand \@sanitize@url [0]{\catcode `\\12\catcode `\$12\catcode
  `\&12\catcode `\#12\catcode `\^12\catcode `\_12\catcode `\%12\relax}%
\providecommand \@@startlink[1]{}%
\providecommand \@@endlink[0]{}%
\providecommand \url  [0]{\begingroup\@sanitize@url \@url }%
\providecommand \@url [1]{\endgroup\@href {#1}{\urlprefix }}%
\providecommand \urlprefix  [0]{URL }%
\providecommand \Eprint [0]{\href }%
\providecommand \doibase [0]{https://doi.org/}%
\providecommand \selectlanguage [0]{\@gobble}%
\providecommand \bibinfo  [0]{\@secondoftwo}%
\providecommand \bibfield  [0]{\@secondoftwo}%
\providecommand \translation [1]{[#1]}%
\providecommand \BibitemOpen [0]{}%
\providecommand \bibitemStop [0]{}%
\providecommand \bibitemNoStop [0]{.\EOS\space}%
\providecommand \EOS [0]{\spacefactor3000\relax}%
\providecommand \BibitemShut  [1]{\csname bibitem#1\endcsname}%
\let\auto@bib@innerbib\@empty
\bibitem [{\citenamefont {Dalela}, \citenamefont {Balaji},\ and\ \citenamefont
  {Jena}(2022)}]{dalela2022review}%
  \BibitemOpen
  \bibfield  {author} {\bibinfo {author} {\bibfnamefont {S.}~\bibnamefont
  {Dalela}}, \bibinfo {author} {\bibfnamefont {P.}~\bibnamefont {Balaji}},\
  and\ \bibinfo {author} {\bibfnamefont {D.}~\bibnamefont {Jena}},\ }\bibfield
  {title} {\enquote {\bibinfo {title} {A review on application of mechanical
  metamaterials for vibration control},}\ }\href
  {https://doi.org/10.1080/15376494.2021.1892244} {\bibfield  {journal}
  {\bibinfo  {journal} {Mechanics of advanced materials and structures}\
  }\textbf {\bibinfo {volume} {29}},\ \bibinfo {pages} {3237--3262} (\bibinfo
  {year} {2022})}\BibitemShut {NoStop}%
\bibitem [{\citenamefont {Xiao}\ \emph {et~al.}(2020)\citenamefont {Xiao},
  \citenamefont {Wang}, \citenamefont {Liu}, \citenamefont {Zhou},
  \citenamefont {Jiang},\ and\ \citenamefont {Zhang}}]{xiao2020active}%
  \BibitemOpen
  \bibfield  {author} {\bibinfo {author} {\bibfnamefont {S.}~\bibnamefont
  {Xiao}}, \bibinfo {author} {\bibfnamefont {T.}~\bibnamefont {Wang}}, \bibinfo
  {author} {\bibfnamefont {T.}~\bibnamefont {Liu}}, \bibinfo {author}
  {\bibfnamefont {C.}~\bibnamefont {Zhou}}, \bibinfo {author} {\bibfnamefont
  {X.}~\bibnamefont {Jiang}},\ and\ \bibinfo {author} {\bibfnamefont
  {J.}~\bibnamefont {Zhang}},\ }\bibfield  {title} {\enquote {\bibinfo {title}
  {Active metamaterials and metadevices: a review},}\ }\href
  {https://doi.org/10.1088/1361-6463/abaced} {\bibfield  {journal} {\bibinfo
  {journal} {Journal of Physics D: Applied Physics}\ }\textbf {\bibinfo
  {volume} {53}},\ \bibinfo {pages} {503002} (\bibinfo {year}
  {2020})}\BibitemShut {NoStop}%
\bibitem [{\citenamefont {Achaoui}\ \emph {et~al.}(2011)\citenamefont
  {Achaoui}, \citenamefont {Khelif}, \citenamefont {Benchabane}, \citenamefont
  {Robert},\ and\ \citenamefont {Laude}}]{achaoui2011experimental}%
  \BibitemOpen
  \bibfield  {author} {\bibinfo {author} {\bibfnamefont {Y.}~\bibnamefont
  {Achaoui}}, \bibinfo {author} {\bibfnamefont {A.}~\bibnamefont {Khelif}},
  \bibinfo {author} {\bibfnamefont {S.}~\bibnamefont {Benchabane}}, \bibinfo
  {author} {\bibfnamefont {L.}~\bibnamefont {Robert}},\ and\ \bibinfo {author}
  {\bibfnamefont {V.}~\bibnamefont {Laude}},\ }\bibfield  {title} {\enquote
  {\bibinfo {title} {Experimental observation of locally-resonant and bragg
  band gaps for surface guided waves in a phononic crystal of pillars},}\
  }\href {https://doi.org/10.1103/PhysRevB.83.104201} {\bibfield  {journal}
  {\bibinfo  {journal} {Physical Review B}\ }\textbf {\bibinfo {volume} {83}},\
  \bibinfo {pages} {104201} (\bibinfo {year} {2011})}\BibitemShut {NoStop}%
\bibitem [{\citenamefont {Kadic}\ \emph {et~al.}(2013)\citenamefont {Kadic},
  \citenamefont {B{\"u}ckmann}, \citenamefont {Schittny},\ and\ \citenamefont
  {Wegener}}]{kadic2013metamaterials}%
  \BibitemOpen
  \bibfield  {author} {\bibinfo {author} {\bibfnamefont {M.}~\bibnamefont
  {Kadic}}, \bibinfo {author} {\bibfnamefont {T.}~\bibnamefont {B{\"u}ckmann}},
  \bibinfo {author} {\bibfnamefont {R.}~\bibnamefont {Schittny}},\ and\
  \bibinfo {author} {\bibfnamefont {M.}~\bibnamefont {Wegener}},\ }\bibfield
  {title} {\enquote {\bibinfo {title} {Metamaterials beyond
  electromagnetism},}\ }\href {https://doi.org/10.1088/0034-4885/76/12/126501}
  {\bibfield  {journal} {\bibinfo  {journal} {Reports on Progress in physics}\
  }\textbf {\bibinfo {volume} {76}},\ \bibinfo {pages} {126501} (\bibinfo
  {year} {2013})}\BibitemShut {NoStop}%
\bibitem [{\citenamefont {Lakes}(2017)}]{lakes2017negative}%
  \BibitemOpen
  \bibfield  {author} {\bibinfo {author} {\bibfnamefont {R.~S.}\ \bibnamefont
  {Lakes}},\ }\bibfield  {title} {\enquote {\bibinfo {title}
  {Negative-poisson's-ratio materials: auxetic solids},}\ }\href
  {https://doi.org/10.1146/annurev-matsci-070616-124118} {\bibfield  {journal}
  {\bibinfo  {journal} {Annual review of materials research}\ }\textbf
  {\bibinfo {volume} {47}},\ \bibinfo {pages} {63--81} (\bibinfo {year}
  {2017})}\BibitemShut {NoStop}%
\bibitem [{\citenamefont {Bertoldi}\ \emph {et~al.}(2017)\citenamefont
  {Bertoldi}, \citenamefont {Vitelli}, \citenamefont {Christensen},\ and\
  \citenamefont {Van~Hecke}}]{bertoldi2017flexible}%
  \BibitemOpen
  \bibfield  {author} {\bibinfo {author} {\bibfnamefont {K.}~\bibnamefont
  {Bertoldi}}, \bibinfo {author} {\bibfnamefont {V.}~\bibnamefont {Vitelli}},
  \bibinfo {author} {\bibfnamefont {J.}~\bibnamefont {Christensen}},\ and\
  \bibinfo {author} {\bibfnamefont {M.}~\bibnamefont {Van~Hecke}},\ }\bibfield
  {title} {\enquote {\bibinfo {title} {Flexible mechanical metamaterials},}\
  }\href {https://doi.org/10.1038/natrevmats.2017.66} {\bibfield  {journal}
  {\bibinfo  {journal} {Nature Reviews Materials}\ }\textbf {\bibinfo {volume}
  {2}},\ \bibinfo {pages} {1--11} (\bibinfo {year} {2017})}\BibitemShut
  {NoStop}%
\bibitem [{\citenamefont {Frenzel}, \citenamefont {Kadic},\ and\ \citenamefont
  {Wegener}(2017)}]{frenzel2017three}%
  \BibitemOpen
  \bibfield  {author} {\bibinfo {author} {\bibfnamefont {T.}~\bibnamefont
  {Frenzel}}, \bibinfo {author} {\bibfnamefont {M.}~\bibnamefont {Kadic}},\
  and\ \bibinfo {author} {\bibfnamefont {M.}~\bibnamefont {Wegener}},\
  }\bibfield  {title} {\enquote {\bibinfo {title} {Three-dimensional mechanical
  metamaterials with a twist},}\ }\href
  {https://doi.org/10.1126/science.aao4640} {\bibfield  {journal} {\bibinfo
  {journal} {Science}\ }\textbf {\bibinfo {volume} {358}},\ \bibinfo {pages}
  {1072--1074} (\bibinfo {year} {2017})}\BibitemShut {NoStop}%
\bibitem [{\citenamefont {Zhong}\ \emph {et~al.}(2019)\citenamefont {Zhong},
  \citenamefont {Fu}, \citenamefont {Chen}, \citenamefont {Zheng},\ and\
  \citenamefont {Hu}}]{zhong2019novel}%
  \BibitemOpen
  \bibfield  {author} {\bibinfo {author} {\bibfnamefont {R.}~\bibnamefont
  {Zhong}}, \bibinfo {author} {\bibfnamefont {M.}~\bibnamefont {Fu}}, \bibinfo
  {author} {\bibfnamefont {X.}~\bibnamefont {Chen}}, \bibinfo {author}
  {\bibfnamefont {B.}~\bibnamefont {Zheng}},\ and\ \bibinfo {author}
  {\bibfnamefont {L.}~\bibnamefont {Hu}},\ }\bibfield  {title} {\enquote
  {\bibinfo {title} {A novel three-dimensional mechanical metamaterial with
  compression-torsion properties},}\ }\href
  {https://doi.org/10.1016/j.compstruct.2019.111232} {\bibfield  {journal}
  {\bibinfo  {journal} {Composite Structures}\ }\textbf {\bibinfo {volume}
  {226}},\ \bibinfo {pages} {111232} (\bibinfo {year} {2019})}\BibitemShut
  {NoStop}%
\bibitem [{\citenamefont {Ungureanu}\ \emph {et~al.}(2015)\citenamefont
  {Ungureanu}, \citenamefont {Achaoui}, \citenamefont {Enoch}, \citenamefont
  {Br{\^u}l{\'e}},\ and\ \citenamefont {Guenneau}}]{ungureanu2015auxetic}%
  \BibitemOpen
  \bibfield  {author} {\bibinfo {author} {\bibfnamefont {B.}~\bibnamefont
  {Ungureanu}}, \bibinfo {author} {\bibfnamefont {Y.}~\bibnamefont {Achaoui}},
  \bibinfo {author} {\bibfnamefont {S.}~\bibnamefont {Enoch}}, \bibinfo
  {author} {\bibfnamefont {S.}~\bibnamefont {Br{\^u}l{\'e}}},\ and\ \bibinfo
  {author} {\bibfnamefont {S.}~\bibnamefont {Guenneau}},\ }\bibfield  {title}
  {\enquote {\bibinfo {title} {Auxetic-like metamaterials as novel earthquake
  protections},}\ }\href {https://doi.org/10.48550/arXiv.1510.08785} {\bibfield
   {journal} {\bibinfo  {journal} {arXiv preprint arXiv:1510.08785}\ }
  (\bibinfo {year} {2015}),\ 10.48550/arXiv.1510.08785}\BibitemShut {NoStop}%
\bibitem [{\citenamefont {Frenzel}\ \emph {et~al.}(2019)\citenamefont
  {Frenzel}, \citenamefont {K{\"o}pfler}, \citenamefont {Jung}, \citenamefont
  {Kadic},\ and\ \citenamefont {Wegener}}]{frenzel2019ultrasound}%
  \BibitemOpen
  \bibfield  {author} {\bibinfo {author} {\bibfnamefont {T.}~\bibnamefont
  {Frenzel}}, \bibinfo {author} {\bibfnamefont {J.}~\bibnamefont
  {K{\"o}pfler}}, \bibinfo {author} {\bibfnamefont {E.}~\bibnamefont {Jung}},
  \bibinfo {author} {\bibfnamefont {M.}~\bibnamefont {Kadic}},\ and\ \bibinfo
  {author} {\bibfnamefont {M.}~\bibnamefont {Wegener}},\ }\bibfield  {title}
  {\enquote {\bibinfo {title} {Ultrasound experiments on acoustical activity in
  chiral mechanical metamaterials},}\ }\href
  {https://doi.org/10.1038/s41467-019-11366-8} {\bibfield  {journal} {\bibinfo
  {journal} {Nature communications}\ }\textbf {\bibinfo {volume} {10}},\
  \bibinfo {pages} {1--6} (\bibinfo {year} {2019})}\BibitemShut {NoStop}%
\bibitem [{\citenamefont {Lemkalli}\ \emph {et~al.}(2022)\citenamefont
  {Lemkalli}, \citenamefont {Kadic}, \citenamefont {Badri}, \citenamefont
  {Guenneau}, \citenamefont {Mir},\ and\ \citenamefont
  {Achaoui}}]{lemkalli2022longitudinal}%
  \BibitemOpen
  \bibfield  {author} {\bibinfo {author} {\bibfnamefont {B.}~\bibnamefont
  {Lemkalli}}, \bibinfo {author} {\bibfnamefont {M.}~\bibnamefont {Kadic}},
  \bibinfo {author} {\bibfnamefont {Y.~E.}\ \bibnamefont {Badri}}, \bibinfo
  {author} {\bibfnamefont {S.}~\bibnamefont {Guenneau}}, \bibinfo {author}
  {\bibfnamefont {A.}~\bibnamefont {Mir}},\ and\ \bibinfo {author}
  {\bibfnamefont {Y.}~\bibnamefont {Achaoui}},\ }\bibfield  {title} {\enquote
  {\bibinfo {title} {Longitudinal-twist wave converter based on chiral
  metamaterials},}\ }\href {https://doi.org/10.48550/arXiv.2211.03222}
  {\bibfield  {journal} {\bibinfo  {journal} {arXiv preprint arXiv:2211.03222}\
  } (\bibinfo {year} {2022}),\ 10.48550/arXiv.2211.03222}\BibitemShut {NoStop}%
\bibitem [{\citenamefont {Xu}\ \emph {et~al.}(2022)\citenamefont {Xu},
  \citenamefont {Wang}, \citenamefont {Tachi},\ and\ \citenamefont
  {Chuang}}]{xu2022origami}%
  \BibitemOpen
  \bibfield  {author} {\bibinfo {author} {\bibfnamefont {Z.-L.}\ \bibnamefont
  {Xu}}, \bibinfo {author} {\bibfnamefont {D.-F.}\ \bibnamefont {Wang}},
  \bibinfo {author} {\bibfnamefont {T.}~\bibnamefont {Tachi}},\ and\ \bibinfo
  {author} {\bibfnamefont {K.-C.}\ \bibnamefont {Chuang}},\ }\bibfield  {title}
  {\enquote {\bibinfo {title} {An origami longitudinal--torsional wave
  converter},}\ }\href {https://doi.org/10.1016/j.eml.2021.101570} {\bibfield
  {journal} {\bibinfo  {journal} {Extreme Mechanics Letters}\ }\textbf
  {\bibinfo {volume} {51}},\ \bibinfo {pages} {101570} (\bibinfo {year}
  {2022})}\BibitemShut {NoStop}%
\bibitem [{\citenamefont {Fano}(1961)}]{fano1961effects}%
  \BibitemOpen
  \bibfield  {author} {\bibinfo {author} {\bibfnamefont {U.}~\bibnamefont
  {Fano}},\ }\bibfield  {title} {\enquote {\bibinfo {title} {Effects of
  configuration interaction on intensities and phase shifts},}\ }\href
  {https://doi.org/10.1103/PhysRev.124.1866} {\bibfield  {journal} {\bibinfo
  {journal} {Physical Review}\ }\textbf {\bibinfo {volume} {124}},\ \bibinfo
  {pages} {1866} (\bibinfo {year} {1961})}\BibitemShut {NoStop}%
\bibitem [{\citenamefont {Zhou}\ \emph {et~al.}(2014)\citenamefont {Zhou},
  \citenamefont {Zhao}, \citenamefont {Shuai}, \citenamefont {Yang},
  \citenamefont {Chuwongin}, \citenamefont {Chadha}, \citenamefont {Seo},
  \citenamefont {Wang}, \citenamefont {Liu}, \citenamefont {Ma} \emph
  {et~al.}}]{zhou2014progress}%
  \BibitemOpen
  \bibfield  {author} {\bibinfo {author} {\bibfnamefont {W.}~\bibnamefont
  {Zhou}}, \bibinfo {author} {\bibfnamefont {D.}~\bibnamefont {Zhao}}, \bibinfo
  {author} {\bibfnamefont {Y.-C.}\ \bibnamefont {Shuai}}, \bibinfo {author}
  {\bibfnamefont {H.}~\bibnamefont {Yang}}, \bibinfo {author} {\bibfnamefont
  {S.}~\bibnamefont {Chuwongin}}, \bibinfo {author} {\bibfnamefont
  {A.}~\bibnamefont {Chadha}}, \bibinfo {author} {\bibfnamefont {J.-H.}\
  \bibnamefont {Seo}}, \bibinfo {author} {\bibfnamefont {K.~X.}\ \bibnamefont
  {Wang}}, \bibinfo {author} {\bibfnamefont {V.}~\bibnamefont {Liu}}, \bibinfo
  {author} {\bibfnamefont {Z.}~\bibnamefont {Ma}}, \emph {et~al.},\ }\bibfield
  {title} {\enquote {\bibinfo {title} {Progress in 2d photonic crystal fano
  resonance photonics},}\ }\href
  {https://doi.org/10.1016/j.pquantelec.2014.01.001} {\bibfield  {journal}
  {\bibinfo  {journal} {Progress in Quantum Electronics}\ }\textbf {\bibinfo
  {volume} {38}},\ \bibinfo {pages} {1--74} (\bibinfo {year}
  {2014})}\BibitemShut {NoStop}%
\bibitem [{\citenamefont {Shuai}\ \emph {et~al.}(2013)\citenamefont {Shuai},
  \citenamefont {Zhao}, \citenamefont {Tian}, \citenamefont {Seo},
  \citenamefont {Plant}, \citenamefont {Ma}, \citenamefont {Fan},\ and\
  \citenamefont {Zhou}}]{shuai2013double}%
  \BibitemOpen
  \bibfield  {author} {\bibinfo {author} {\bibfnamefont {Y.}~\bibnamefont
  {Shuai}}, \bibinfo {author} {\bibfnamefont {D.}~\bibnamefont {Zhao}},
  \bibinfo {author} {\bibfnamefont {Z.}~\bibnamefont {Tian}}, \bibinfo {author}
  {\bibfnamefont {J.-H.}\ \bibnamefont {Seo}}, \bibinfo {author} {\bibfnamefont
  {D.~V.}\ \bibnamefont {Plant}}, \bibinfo {author} {\bibfnamefont
  {Z.}~\bibnamefont {Ma}}, \bibinfo {author} {\bibfnamefont {S.}~\bibnamefont
  {Fan}},\ and\ \bibinfo {author} {\bibfnamefont {W.}~\bibnamefont {Zhou}},\
  }\bibfield  {title} {\enquote {\bibinfo {title} {Double-layer fano resonance
  photonic crystal filters},}\ }\href {https://doi.org/10.1364/OE.21.024582}
  {\bibfield  {journal} {\bibinfo  {journal} {Optics Express}\ }\textbf
  {\bibinfo {volume} {21}},\ \bibinfo {pages} {24582--24589} (\bibinfo {year}
  {2013})}\BibitemShut {NoStop}%
\bibitem [{\citenamefont {Luk'Yanchuk}\ \emph {et~al.}(2010)\citenamefont
  {Luk'Yanchuk}, \citenamefont {Zheludev}, \citenamefont {Maier}, \citenamefont
  {Halas}, \citenamefont {Nordlander}, \citenamefont {Giessen},\ and\
  \citenamefont {Chong}}]{luk2010fano}%
  \BibitemOpen
  \bibfield  {author} {\bibinfo {author} {\bibfnamefont {B.}~\bibnamefont
  {Luk'Yanchuk}}, \bibinfo {author} {\bibfnamefont {N.~I.}\ \bibnamefont
  {Zheludev}}, \bibinfo {author} {\bibfnamefont {S.~A.}\ \bibnamefont {Maier}},
  \bibinfo {author} {\bibfnamefont {N.~J.}\ \bibnamefont {Halas}}, \bibinfo
  {author} {\bibfnamefont {P.}~\bibnamefont {Nordlander}}, \bibinfo {author}
  {\bibfnamefont {H.}~\bibnamefont {Giessen}},\ and\ \bibinfo {author}
  {\bibfnamefont {C.~T.}\ \bibnamefont {Chong}},\ }\bibfield  {title} {\enquote
  {\bibinfo {title} {The fano resonance in plasmonic nanostructures and
  metamaterials},}\ }\href {https://doi.org/10.1038/nmat2810} {\bibfield
  {journal} {\bibinfo  {journal} {Nature materials}\ }\textbf {\bibinfo
  {volume} {9}},\ \bibinfo {pages} {707--715} (\bibinfo {year}
  {2010})}\BibitemShut {NoStop}%
\bibitem [{\citenamefont {Wang}\ \emph {et~al.}(2020)\citenamefont {Wang},
  \citenamefont {Jin}, \citenamefont {Wang}, \citenamefont {Bonello},
  \citenamefont {Djafari-Rouhani},\ and\ \citenamefont
  {Fleury}}]{wang2020robust}%
  \BibitemOpen
  \bibfield  {author} {\bibinfo {author} {\bibfnamefont {W.}~\bibnamefont
  {Wang}}, \bibinfo {author} {\bibfnamefont {Y.}~\bibnamefont {Jin}}, \bibinfo
  {author} {\bibfnamefont {W.}~\bibnamefont {Wang}}, \bibinfo {author}
  {\bibfnamefont {B.}~\bibnamefont {Bonello}}, \bibinfo {author} {\bibfnamefont
  {B.}~\bibnamefont {Djafari-Rouhani}},\ and\ \bibinfo {author} {\bibfnamefont
  {R.}~\bibnamefont {Fleury}},\ }\bibfield  {title} {\enquote {\bibinfo {title}
  {Robust fano resonance in a topological mechanical beam},}\ }\href
  {https://doi.org/10.1103/PhysRevB.101.024101} {\bibfield  {journal} {\bibinfo
   {journal} {Physical Review B}\ }\textbf {\bibinfo {volume} {101}},\ \bibinfo
  {pages} {024101} (\bibinfo {year} {2020})}\BibitemShut {NoStop}%
\bibitem [{\citenamefont {El~Boudouti}\ \emph {et~al.}(2008)\citenamefont
  {El~Boudouti}, \citenamefont {Mrabti}, \citenamefont {Al-Wahsh},
  \citenamefont {Djafari-Rouhani}, \citenamefont {Akjouj},\ and\ \citenamefont
  {Dobrzynski}}]{el2008transmission}%
  \BibitemOpen
  \bibfield  {author} {\bibinfo {author} {\bibfnamefont {E.}~\bibnamefont
  {El~Boudouti}}, \bibinfo {author} {\bibfnamefont {T.}~\bibnamefont {Mrabti}},
  \bibinfo {author} {\bibfnamefont {H.}~\bibnamefont {Al-Wahsh}}, \bibinfo
  {author} {\bibfnamefont {B.}~\bibnamefont {Djafari-Rouhani}}, \bibinfo
  {author} {\bibfnamefont {A.}~\bibnamefont {Akjouj}},\ and\ \bibinfo {author}
  {\bibfnamefont {L.}~\bibnamefont {Dobrzynski}},\ }\bibfield  {title}
  {\enquote {\bibinfo {title} {Transmission gaps and fano resonances in an
  acoustic waveguide: analytical model},}\ }\href
  {https://doi.org/10.1088/0953-8984/20/25/255212} {\bibfield  {journal}
  {\bibinfo  {journal} {Journal of Physics: Condensed Matter}\ }\textbf
  {\bibinfo {volume} {20}},\ \bibinfo {pages} {255212} (\bibinfo {year}
  {2008})}\BibitemShut {NoStop}%
\bibitem [{\citenamefont {Amin}\ \emph {et~al.}(2015)\citenamefont {Amin},
  \citenamefont {Elayouch}, \citenamefont {Farhat}, \citenamefont {Addouche},
  \citenamefont {Khelif},\ and\ \citenamefont
  {Ba{\u{g}}c{\i}}}]{amin2015acoustically}%
  \BibitemOpen
  \bibfield  {author} {\bibinfo {author} {\bibfnamefont {M.}~\bibnamefont
  {Amin}}, \bibinfo {author} {\bibfnamefont {A.}~\bibnamefont {Elayouch}},
  \bibinfo {author} {\bibfnamefont {M.}~\bibnamefont {Farhat}}, \bibinfo
  {author} {\bibfnamefont {M.}~\bibnamefont {Addouche}}, \bibinfo {author}
  {\bibfnamefont {A.}~\bibnamefont {Khelif}},\ and\ \bibinfo {author}
  {\bibfnamefont {H.}~\bibnamefont {Ba{\u{g}}c{\i}}},\ }\bibfield  {title}
  {\enquote {\bibinfo {title} {Acoustically induced transparency using fano
  resonant periodic arrays},}\ }\href {https://doi.org/10.1063/1.4934247}
  {\bibfield  {journal} {\bibinfo  {journal} {Journal of Applied Physics}\
  }\textbf {\bibinfo {volume} {118}},\ \bibinfo {pages} {164901} (\bibinfo
  {year} {2015})}\BibitemShut {NoStop}%
\bibitem [{\citenamefont {Qi}\ \emph {et~al.}(2014)\citenamefont {Qi},
  \citenamefont {Yu}, \citenamefont {Wang}, \citenamefont {Wang},\ and\
  \citenamefont {Wang}}]{qi2014interference}%
  \BibitemOpen
  \bibfield  {author} {\bibinfo {author} {\bibfnamefont {L.}~\bibnamefont
  {Qi}}, \bibinfo {author} {\bibfnamefont {G.}~\bibnamefont {Yu}}, \bibinfo
  {author} {\bibfnamefont {X.}~\bibnamefont {Wang}}, \bibinfo {author}
  {\bibfnamefont {G.}~\bibnamefont {Wang}},\ and\ \bibinfo {author}
  {\bibfnamefont {N.}~\bibnamefont {Wang}},\ }\bibfield  {title} {\enquote
  {\bibinfo {title} {Interference-induced angle-independent acoustical
  transparency},}\ }\href {https://doi.org/10.1063/1.4904525} {\bibfield
  {journal} {\bibinfo  {journal} {Journal of Applied Physics}\ }\textbf
  {\bibinfo {volume} {116}},\ \bibinfo {pages} {234506} (\bibinfo {year}
  {2014})}\BibitemShut {NoStop}%
\bibitem [{\citenamefont {Zaki}\ \emph {et~al.}(2020)\citenamefont {Zaki},
  \citenamefont {Mehaney}, \citenamefont {Hassanein},\ and\ \citenamefont
  {Aly}}]{zaki2020fano}%
  \BibitemOpen
  \bibfield  {author} {\bibinfo {author} {\bibfnamefont {S.~E.}\ \bibnamefont
  {Zaki}}, \bibinfo {author} {\bibfnamefont {A.}~\bibnamefont {Mehaney}},
  \bibinfo {author} {\bibfnamefont {H.~M.}\ \bibnamefont {Hassanein}},\ and\
  \bibinfo {author} {\bibfnamefont {A.~H.}\ \bibnamefont {Aly}},\ }\bibfield
  {title} {\enquote {\bibinfo {title} {Fano resonance based defected 1d
  phononic crystal for highly sensitive gas sensing applications},}\ }\href
  {https://doi.org/10.1038/s41598-020-75076-8} {\bibfield  {journal} {\bibinfo
  {journal} {Scientific Reports}\ }\textbf {\bibinfo {volume} {10}},\ \bibinfo
  {pages} {1--16} (\bibinfo {year} {2020})}\BibitemShut {NoStop}%
\bibitem [{\citenamefont {Goffaux}\ \emph {et~al.}(2002)\citenamefont
  {Goffaux}, \citenamefont {S{\'a}nchez-Dehesa}, \citenamefont {Yeyati},
  \citenamefont {Lambin}, \citenamefont {Khelif}, \citenamefont {Vasseur},\
  and\ \citenamefont {Djafari-Rouhani}}]{goffaux2002evidence}%
  \BibitemOpen
  \bibfield  {author} {\bibinfo {author} {\bibfnamefont {C.}~\bibnamefont
  {Goffaux}}, \bibinfo {author} {\bibfnamefont {J.}~\bibnamefont
  {S{\'a}nchez-Dehesa}}, \bibinfo {author} {\bibfnamefont {A.~L.}\ \bibnamefont
  {Yeyati}}, \bibinfo {author} {\bibfnamefont {P.}~\bibnamefont {Lambin}},
  \bibinfo {author} {\bibfnamefont {A.}~\bibnamefont {Khelif}}, \bibinfo
  {author} {\bibfnamefont {J.}~\bibnamefont {Vasseur}},\ and\ \bibinfo {author}
  {\bibfnamefont {B.}~\bibnamefont {Djafari-Rouhani}},\ }\bibfield  {title}
  {\enquote {\bibinfo {title} {Evidence of fano-like interference phenomena in
  locally resonant materials},}\ }\href
  {https://doi.org/10.1103/PhysRevLett.88.225502} {\bibfield  {journal}
  {\bibinfo  {journal} {Physical review letters}\ }\textbf {\bibinfo {volume}
  {88}},\ \bibinfo {pages} {225502} (\bibinfo {year} {2002})}\BibitemShut
  {NoStop}%
\bibitem [{\citenamefont {Oudich}\ \emph {et~al.}(2018)\citenamefont {Oudich},
  \citenamefont {Djafari-Rouhani}, \citenamefont {Bonello}, \citenamefont
  {Pennec}, \citenamefont {Hemaidia}, \citenamefont {Sarry},\ and\
  \citenamefont {Beyssen}}]{oudich2018rayleigh}%
  \BibitemOpen
  \bibfield  {author} {\bibinfo {author} {\bibfnamefont {M.}~\bibnamefont
  {Oudich}}, \bibinfo {author} {\bibfnamefont {B.}~\bibnamefont
  {Djafari-Rouhani}}, \bibinfo {author} {\bibfnamefont {B.}~\bibnamefont
  {Bonello}}, \bibinfo {author} {\bibfnamefont {Y.}~\bibnamefont {Pennec}},
  \bibinfo {author} {\bibfnamefont {S.}~\bibnamefont {Hemaidia}}, \bibinfo
  {author} {\bibfnamefont {F.}~\bibnamefont {Sarry}},\ and\ \bibinfo {author}
  {\bibfnamefont {D.}~\bibnamefont {Beyssen}},\ }\bibfield  {title} {\enquote
  {\bibinfo {title} {Rayleigh waves in phononic crystal made of multilayered
  pillars: confined modes, fano resonances, and acoustically induced
  transparency},}\ }\href {https://doi.org/10.1103/PhysRevApplied.9.034013}
  {\bibfield  {journal} {\bibinfo  {journal} {Physical Review Applied}\
  }\textbf {\bibinfo {volume} {9}},\ \bibinfo {pages} {034013} (\bibinfo {year}
  {2018})}\BibitemShut {NoStop}%
\bibitem [{\citenamefont {Sun}\ \emph {et~al.}(2019)\citenamefont {Sun},
  \citenamefont {Xia}, \citenamefont {Sun}, \citenamefont {Yuan}, \citenamefont
  {Ge},\ and\ \citenamefont {Liu}}]{sun2019dual}%
  \BibitemOpen
  \bibfield  {author} {\bibinfo {author} {\bibfnamefont {Y.-Y.}\ \bibnamefont
  {Sun}}, \bibinfo {author} {\bibfnamefont {J.-P.}\ \bibnamefont {Xia}},
  \bibinfo {author} {\bibfnamefont {H.-X.}\ \bibnamefont {Sun}}, \bibinfo
  {author} {\bibfnamefont {S.-Q.}\ \bibnamefont {Yuan}}, \bibinfo {author}
  {\bibfnamefont {Y.}~\bibnamefont {Ge}},\ and\ \bibinfo {author}
  {\bibfnamefont {X.-J.}\ \bibnamefont {Liu}},\ }\bibfield  {title} {\enquote
  {\bibinfo {title} {Dual-band fano resonance of low-frequency sound based on
  artificial mie resonances},}\ }\href {https://doi.org/10.1002/advs.201901307}
  {\bibfield  {journal} {\bibinfo  {journal} {Advanced Science}\ }\textbf
  {\bibinfo {volume} {6}},\ \bibinfo {pages} {1901307} (\bibinfo {year}
  {2019})}\BibitemShut {NoStop}%
\bibitem [{\citenamefont {Cummer}, \citenamefont {Christensen},\ and\
  \citenamefont {Al{\`u}}(2016)}]{cummer2016controlling}%
  \BibitemOpen
  \bibfield  {author} {\bibinfo {author} {\bibfnamefont {S.~A.}\ \bibnamefont
  {Cummer}}, \bibinfo {author} {\bibfnamefont {J.}~\bibnamefont
  {Christensen}},\ and\ \bibinfo {author} {\bibfnamefont {A.}~\bibnamefont
  {Al{\`u}}},\ }\bibfield  {title} {\enquote {\bibinfo {title} {Controlling
  sound with acoustic metamaterials},}\ }\href
  {https://doi.org/10.1038/natrevmats.2016.1} {\bibfield  {journal} {\bibinfo
  {journal} {Nature Reviews Materials}\ }\textbf {\bibinfo {volume} {1}},\
  \bibinfo {pages} {1--13} (\bibinfo {year} {2016})}\BibitemShut {NoStop}%
\bibitem [{\citenamefont {Bergamini}\ \emph {et~al.}(2019)\citenamefont
  {Bergamini}, \citenamefont {Miniaci}, \citenamefont {Delpero}, \citenamefont
  {Tallarico}, \citenamefont {Van~Damme}, \citenamefont {Hannema},
  \citenamefont {Leibacher},\ and\ \citenamefont
  {Zemp}}]{bergamini2019tacticity}%
  \BibitemOpen
  \bibfield  {author} {\bibinfo {author} {\bibfnamefont {A.}~\bibnamefont
  {Bergamini}}, \bibinfo {author} {\bibfnamefont {M.}~\bibnamefont {Miniaci}},
  \bibinfo {author} {\bibfnamefont {T.}~\bibnamefont {Delpero}}, \bibinfo
  {author} {\bibfnamefont {D.}~\bibnamefont {Tallarico}}, \bibinfo {author}
  {\bibfnamefont {B.}~\bibnamefont {Van~Damme}}, \bibinfo {author}
  {\bibfnamefont {G.}~\bibnamefont {Hannema}}, \bibinfo {author} {\bibfnamefont
  {I.}~\bibnamefont {Leibacher}},\ and\ \bibinfo {author} {\bibfnamefont
  {A.}~\bibnamefont {Zemp}},\ }\bibfield  {title} {\enquote {\bibinfo {title}
  {Tacticity in chiral phononic crystals},}\ }\href
  {https://doi.org/10.1038/s41467-019-12587-7} {\bibfield  {journal} {\bibinfo
  {journal} {Nature communications}\ }\textbf {\bibinfo {volume} {10}},\
  \bibinfo {pages} {1--8} (\bibinfo {year} {2019})}\BibitemShut {NoStop}%
\bibitem [{\citenamefont {Lucklum}, \citenamefont {Ke},\ and\ \citenamefont
  {Zubtsov}(2012)}]{lucklum2012two}%
  \BibitemOpen
  \bibfield  {author} {\bibinfo {author} {\bibfnamefont {R.}~\bibnamefont
  {Lucklum}}, \bibinfo {author} {\bibfnamefont {M.}~\bibnamefont {Ke}},\ and\
  \bibinfo {author} {\bibfnamefont {M.}~\bibnamefont {Zubtsov}},\ }\bibfield
  {title} {\enquote {\bibinfo {title} {Two-dimensional phononic crystal sensor
  based on a cavity mode},}\ }\href {https://doi.org/10.1016/j.snb.2012.03.063}
  {\bibfield  {journal} {\bibinfo  {journal} {Sensors and Actuators B:
  Chemical}\ }\textbf {\bibinfo {volume} {171}},\ \bibinfo {pages} {271--277}
  (\bibinfo {year} {2012})}\BibitemShut {NoStop}%
\bibitem [{\citenamefont {Ke}, \citenamefont {Zubtsov},\ and\ \citenamefont
  {Lucklum}(2011)}]{ke2011sub}%
  \BibitemOpen
  \bibfield  {author} {\bibinfo {author} {\bibfnamefont {M.}~\bibnamefont
  {Ke}}, \bibinfo {author} {\bibfnamefont {M.}~\bibnamefont {Zubtsov}},\ and\
  \bibinfo {author} {\bibfnamefont {R.}~\bibnamefont {Lucklum}},\ }\bibfield
  {title} {\enquote {\bibinfo {title} {Sub-wavelength phononic crystal liquid
  sensor},}\ }\href {https://doi.org/10.1063/1.3610391} {\  (\bibinfo {year}
  {2011}),\ 10.1063/1.3610391}\BibitemShut {NoStop}%
\bibitem [{\citenamefont {Oseev}\ \emph {et~al.}(2018)\citenamefont {Oseev},
  \citenamefont {Mukhin}, \citenamefont {Lucklum}, \citenamefont {Zubtsov},
  \citenamefont {Schmidt}, \citenamefont {Steinmann}, \citenamefont {Fomin},
  \citenamefont {Kozyrev},\ and\ \citenamefont {Hirsch}}]{oseev2018study}%
  \BibitemOpen
  \bibfield  {author} {\bibinfo {author} {\bibfnamefont {A.}~\bibnamefont
  {Oseev}}, \bibinfo {author} {\bibfnamefont {N.}~\bibnamefont {Mukhin}},
  \bibinfo {author} {\bibfnamefont {R.}~\bibnamefont {Lucklum}}, \bibinfo
  {author} {\bibfnamefont {M.}~\bibnamefont {Zubtsov}}, \bibinfo {author}
  {\bibfnamefont {M.-P.}\ \bibnamefont {Schmidt}}, \bibinfo {author}
  {\bibfnamefont {U.}~\bibnamefont {Steinmann}}, \bibinfo {author}
  {\bibfnamefont {A.}~\bibnamefont {Fomin}}, \bibinfo {author} {\bibfnamefont
  {A.}~\bibnamefont {Kozyrev}},\ and\ \bibinfo {author} {\bibfnamefont
  {S.}~\bibnamefont {Hirsch}},\ }\bibfield  {title} {\enquote {\bibinfo {title}
  {Study of liquid resonances in solid-liquid composite periodic structures
  (phononic crystals)--theoretical investigations and practical application for
  in-line analysis of conventional petroleum products},}\ }\href
  {https://doi.org/10.1016/j.snb.2017.10.144} {\bibfield  {journal} {\bibinfo
  {journal} {Sensors and Actuators B: Chemical}\ }\textbf {\bibinfo {volume}
  {257}},\ \bibinfo {pages} {469--477} (\bibinfo {year} {2018})}\BibitemShut
  {NoStop}%
\bibitem [{\citenamefont {Wang}\ \emph {et~al.}(2017)\citenamefont {Wang},
  \citenamefont {Wang}, \citenamefont {Wang},\ and\ \citenamefont
  {Laude}}]{wang2017tunable}%
  \BibitemOpen
  \bibfield  {author} {\bibinfo {author} {\bibfnamefont {T.-T.}\ \bibnamefont
  {Wang}}, \bibinfo {author} {\bibfnamefont {Y.-F.}\ \bibnamefont {Wang}},
  \bibinfo {author} {\bibfnamefont {Y.-S.}\ \bibnamefont {Wang}},\ and\
  \bibinfo {author} {\bibfnamefont {V.}~\bibnamefont {Laude}},\ }\bibfield
  {title} {\enquote {\bibinfo {title} {Tunable fluid-filled phononic
  metastrip},}\ }\href {https://doi.org/10.1063/1.4985167} {\bibfield
  {journal} {\bibinfo  {journal} {Applied Physics Letters}\ }\textbf {\bibinfo
  {volume} {111}},\ \bibinfo {pages} {041906} (\bibinfo {year}
  {2017})}\BibitemShut {NoStop}%
\bibitem [{\citenamefont {Wang}\ \emph {et~al.}(2022)\citenamefont {Wang},
  \citenamefont {Wang}, \citenamefont {Deng}, \citenamefont {Laude},\ and\
  \citenamefont {Wang}}]{wang2022reconfigurable}%
  \BibitemOpen
  \bibfield  {author} {\bibinfo {author} {\bibfnamefont {T.-T.}\ \bibnamefont
  {Wang}}, \bibinfo {author} {\bibfnamefont {Y.-F.}\ \bibnamefont {Wang}},
  \bibinfo {author} {\bibfnamefont {Z.-C.}\ \bibnamefont {Deng}}, \bibinfo
  {author} {\bibfnamefont {V.}~\bibnamefont {Laude}},\ and\ \bibinfo {author}
  {\bibfnamefont {Y.-S.}\ \bibnamefont {Wang}},\ }\bibfield  {title} {\enquote
  {\bibinfo {title} {Reconfigurable waveguides defined by selective fluid
  filling in two-dimensional phononic metaplates},}\ }\href
  {https://doi.org/10.1016/j.ymssp.2021.108392} {\bibfield  {journal} {\bibinfo
   {journal} {Mechanical Systems and Signal Processing}\ }\textbf {\bibinfo
  {volume} {165}},\ \bibinfo {pages} {108392} (\bibinfo {year}
  {2022})}\BibitemShut {NoStop}%
\bibitem [{\citenamefont {Gueddida}\ \emph {et~al.}(2021)\citenamefont
  {Gueddida}, \citenamefont {Pennec}, \citenamefont {Zhang}, \citenamefont
  {Lucklum}, \citenamefont {Vellekoop}, \citenamefont {Mukhin}, \citenamefont
  {Lucklum}, \citenamefont {Bonello},\ and\ \citenamefont
  {Djafari~Rouhani}}]{gueddida2021tubular}%
  \BibitemOpen
  \bibfield  {author} {\bibinfo {author} {\bibfnamefont {A.}~\bibnamefont
  {Gueddida}}, \bibinfo {author} {\bibfnamefont {Y.}~\bibnamefont {Pennec}},
  \bibinfo {author} {\bibfnamefont {V.}~\bibnamefont {Zhang}}, \bibinfo
  {author} {\bibfnamefont {F.}~\bibnamefont {Lucklum}}, \bibinfo {author}
  {\bibfnamefont {M.}~\bibnamefont {Vellekoop}}, \bibinfo {author}
  {\bibfnamefont {N.}~\bibnamefont {Mukhin}}, \bibinfo {author} {\bibfnamefont
  {R.}~\bibnamefont {Lucklum}}, \bibinfo {author} {\bibfnamefont
  {B.}~\bibnamefont {Bonello}},\ and\ \bibinfo {author} {\bibfnamefont
  {B.}~\bibnamefont {Djafari~Rouhani}},\ }\bibfield  {title} {\enquote
  {\bibinfo {title} {Tubular phononic crystal sensor},}\ }\href
  {https://doi.org/10.1063/5.0051660} {\bibfield  {journal} {\bibinfo
  {journal} {Journal of Applied Physics}\ }\textbf {\bibinfo {volume} {130}},\
  \bibinfo {pages} {105103} (\bibinfo {year} {2021})}\BibitemShut {NoStop}%
\bibitem [{\citenamefont {Gueddida}\ \emph {et~al.}(2022)\citenamefont
  {Gueddida}, \citenamefont {Pennec}, \citenamefont {Silveira~Fiates},
  \citenamefont {Vellekoop}, \citenamefont {Bonello},\ and\ \citenamefont
  {Djafari-Rouhani}}]{gueddida2022acoustic}%
  \BibitemOpen
  \bibfield  {author} {\bibinfo {author} {\bibfnamefont {A.}~\bibnamefont
  {Gueddida}}, \bibinfo {author} {\bibfnamefont {Y.}~\bibnamefont {Pennec}},
  \bibinfo {author} {\bibfnamefont {A.~L.}\ \bibnamefont {Silveira~Fiates}},
  \bibinfo {author} {\bibfnamefont {M.~J.}\ \bibnamefont {Vellekoop}}, \bibinfo
  {author} {\bibfnamefont {B.}~\bibnamefont {Bonello}},\ and\ \bibinfo {author}
  {\bibfnamefont {B.}~\bibnamefont {Djafari-Rouhani}},\ }\bibfield  {title}
  {\enquote {\bibinfo {title} {Acoustic sensor based on a cylindrical resonator
  for monitoring a liquid flow},}\ }\href
  {https://doi.org/10.3390/cryst12101398} {\bibfield  {journal} {\bibinfo
  {journal} {Crystals}\ }\textbf {\bibinfo {volume} {12}},\ \bibinfo {pages}
  {1398} (\bibinfo {year} {2022})}\BibitemShut {NoStop}%
\end{thebibliography}%

\end{document}